\documentclass[aip,apl,twocolumn,10pt]{revtex4-1}

\usepackage{xcolor}
\usepackage{graphicx}
\usepackage{dcolumn}
\usepackage{bm}
\usepackage{amsfonts}
\usepackage{amsmath}
\usepackage{enumerate}

\newcommand{\Fig} [1]  {Fig.~\ref{#1}}

\begin{document}

\title{Pyridine intercalated Bi$_2$Se$_3$ heterostructures: controlling the 
	topologically protected states}

\author{I. S. S. de Oliveira}
\email{igor.oliveira@dfi.ufla.br}
\affiliation{Departamento de F\'isica, Universidade Federal de Lavras,
   C.P. 3037, 37200-000, Lavras, MG, Brazil}
\author{R. H. Miwa}
\affiliation{Instituto de F\'isica, Universidade Federal de Uberl\^andia,
        C.P. 593, 38400-902, Uberl\^andia, MG, Brazil}
          
\date{\today}
    
\begin{abstract}

We use \textit{ab initio} simulations to investigate the 
incorporation of pyridine molecules (C$_5$H$_5$N) in the van der 
Waals gaps of Bi$_2$Se$_3$. The intercalated  pyridine molecules 
increase the separation distance between the Bi$_2$Se$_3$ quintuple layers 
(QLs), suppressing the parity inversion of the 
electronic states at the $\Gamma$-point. We find that the intercalated region 
becomes a trivial insulator. By combining the pristine Bi$_2$Se$_3$  region with 
the one intercalated by the molecules, we have a non-trivial/trivial 
heterojunction characterized by the presence of  (topologically protected) 
metallic  states at the interfacial region. Next we apply  an 
external compressive pressure to the system, and the results are (i) a decrease  
on the separation distance between the QLs intercalated by pyridine molecules, 
and (ii) the metallic states are shifted toward the bulk region, turning the 
system back to insulator. That is, through a suitable  tuning 
of  the external pressure in Bi$_2$Se$_3$, intercalated by pyridine molecules, 
we  can control its topological properties; turning-on and -off the 
topologically protected metallic states  lying at the non-trivial/trivial 
interface.
\end{abstract}

\maketitle

Three dimensional topological insulators (TIs) are insulator materials in the 
bulk phase, but they present metallic topological  surface states (TSSs), 
which are protected  by time-reversal symmetry; as a result 
backscattering processes by time-reversal invariant impurities or defects are 
forbidden \cite{RevModPhys.82.3045}. These materials are of great promise for 
spintronics applications, due to the formation of nearly 
dissipationless spin-polarized surface current \cite{NatMat.11.409.2012}. 
Currently, Bi$_2$Se$_3$ is one of the most investigated TI due to its large band 
gap ($\sim$~0.3~eV), nearly idealized single Dirac cone and for being a pure 
compound \cite{NatPhys.5.438, NatPhys.5.398, NatPhys.460.1101}. Bi$_2$Se$_3$ 
presents a rhombohedral structure composed by quintuple layers (QLs) of Se and 
Bi atoms, forming a sequence of Se--Bi--Se--Bi--Se atoms covalently bonded; 
these QLs are stacked along the $c$--axis of a hexagonal structure by van der 
Waals (vdW) interactions. 

Recent experiments have been exploring the possibility of inserting guest 
species in the vdW gaps of Bi$_2$Se$_3$, this process is known as intercalation. 
Koski {\it et al.}\,\cite{doi:10.1021/ja304925t,doi:10.1021/ja300368x} have 
added various zerovalent metals in the vdW gaps of  Bi$_2$Se$_3$ nanoribbons, it 
is expected that new properties and/or tuning of the Bi$_2$Se$_3$ properties can 
be achieved by intercalation, e.g. superconductivity 
\cite{PhysRevLett.104.057001}. Beside metals, molecules have 
been intercalated in Bi$_2$Se$_3$, for instance, it has been 
shown that pyridine molecules in Bi$_2$Se$_3$ present interesting properties for 
 optoelectronic applications\,\cite{doi:10.1021/nl402937g}. 

In this work we aim to investigate the geometry and electronic structure of 
Bi$_2$Se$_3$ intercalated by  pyridine molecules (py-Bi$_2$Se$_3$).  Upon the 
presence of the  pyridine molecules between the  QLs, we find that the 
py-Bi$_2$Se$_3$ system becomes a  trivial insulator. By considering a 
heterojunction composed by  py-Bi$_2$Se$_3$ and pristine Bi$_2$Se$_3$ 
(py-Bi$_2$Se$_3$/Bi$_2$Se$_3$),  we analyze (i) the occurrence of topologically 
protected metallic states embedded in py-Bi$_2$Se$_3$/Bi$_2$Se$_3$, and (ii) the 
control of those metallic states (turning-on and -off) upon  the application of 
external pressure. 

The calculations are performed based on the density-functional theory (DFT) as 
implemented in the Vienna \textit{ab initio} simulation package (VASP) 
\cite{Kresse}. We use the generalized gradient approximation (GGA), in the form 
proposed by Perdew, Burke and Ernzerhof \cite{PBE}, to describe the 
exchange-correlation functional.  The Kohn-Sham orbitals are expanded in a plane 
wave basis set with an energy cutoff of 400~eV. The electron-ion interactions 
are taken into account using the Projector Augmented Wave (PAW) method 
\cite{Kresse99}. All geometries have been relaxed until atomic forces were lower 
than 0.02~eV/\AA. The  Brillouin Zone is sampled according to the Monkhorst-Pack 
method \cite{Monkhorst}, using at least a 
3$\times$3$\times$1 mesh. We have also 
used a functional that accounts for dispersion effects, representing van der 
Waals (vdW) forces, according to the method developed by Tkatchenko-Scheffler 
(TS)~\cite{Tkatchenko}, which is implemented on VASP \cite{Bucko}. The inclusion 
of van der Waals forces in the simulations is necessary to obtain the correct 
vdW gap between two QLs \cite{PhysRevB.72.184101}, the interaction between 
molecule and Bi$_2$Se$_3$ is also better described with the  inclusion of vdW 
interactions. 

\begin{figure}
  \centering
  \includegraphics[clip,width=7.8cm]{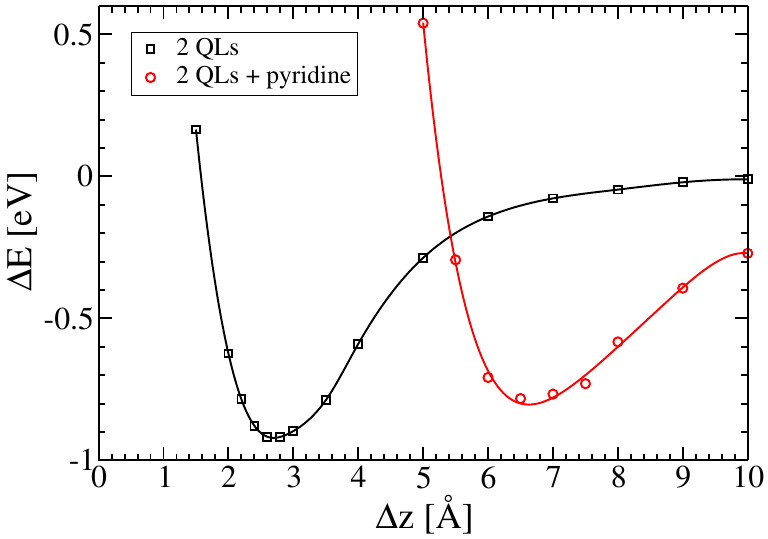}
  \caption{\label{Ef_2QL}  Formation energy for two Bi$_2$Se$_3$ QLs at various
   separation distances. Black line (circles) represents the pristine system and red line
    (squares) the system with a pyridine molecule intercalated between the 2 QLs.}
\end{figure}

Initially we calculate the equilibrium distance ($z_0$) between 
two consecutive QLs of Bi$_2$Se$_3$, vdW gap. Here we  have considered two 
isolated QLs  and minimize the total 
energy as a function of the width of vdW gap ($\Delta z$)\,\cite{relax}. As 
shown in \Fig{Ef_2QL} (squares), we find the energy minimum  for $z_0$ = 
2.65\,\AA; this result is in agreement with experimental 
measurements\,\cite{Lind200347}, and  recent theoretical  
studies\,\cite{PhysRevB.72.184101,NewJPhys.32.7979}. By using the  same 
procedure, we determine $z_0$ upon the presence of a pyridine molecule 
intercalated in between the QLs.  The QLs present a 2$\times$2 surface 
periodicity, which corresponds to a molecular concentration of 
1.64$\times$10$^{-2}$~mol./\AA$^2$. 
In this case, we find that (i)  there is no 
chemical bonding between the molecule and the Bi$_2$Se$_3$ QLs,  and (ii)  the 
vdW gap increases by 3.85\,\AA, $z_0$ = 2.65\,$\rightarrow$\,6.50\,\AA, 
\Fig{Ef_2QL} (circles). Such an increase of the vdW gap reduces the binding 
energy between the QLs, however, the formation of pyridine intercalated 
Bi$_2$Se$_3$ QLs is still an exothermic process.

\begin{figure}
  \centering
  \includegraphics[clip,width=8.5cm]{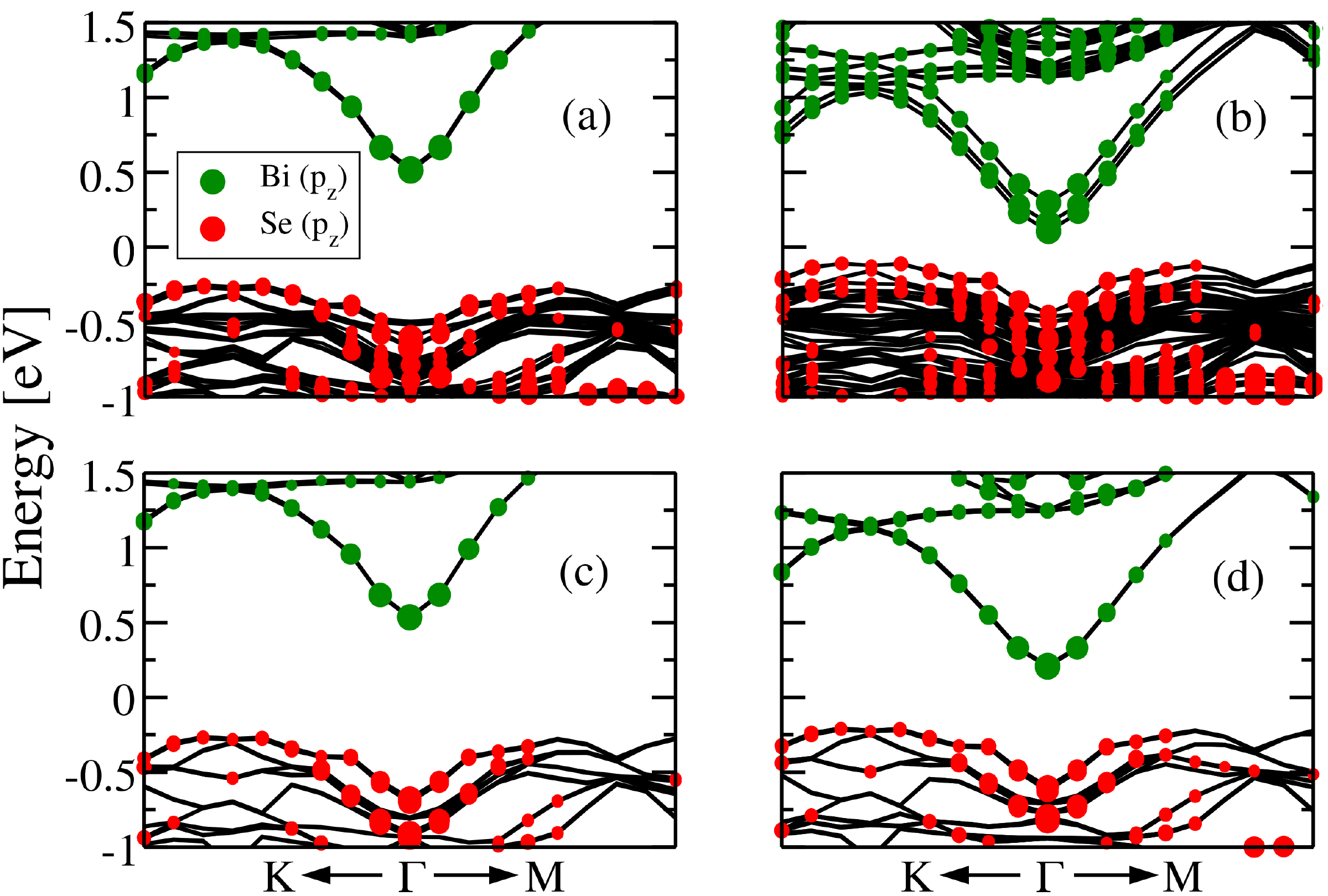}
  \caption{\label{bulk3mol} Band structures for 
  (a) Bi$_2$Se$_3$ bulk
   with intercalated pyridine molecules excluding SOC, and in 
   (b) including SOC.
   (c) Bi$_2$Se$_3$ bulk with the vdW gap 
   obtained in the system with intercalated molecules (\Fig{Ef_2QL}), 
   excluding SOC; and in 
   (d) including SOC.      
   The symbols are contributions of $pz$--orbitals for Se (red) and Bi (green) atoms.}
\end{figure}

TIs present a parity inversion between valence and conduction 
bands due to spin--orbit coupling (SOC). For Bi$_2$Se$_3$ the bulk phase 
presents a band parity inversion at the $\Gamma$-point, 
 namely the  SOC induces a 
band inversion between Se $pz$-orbitals in the valence band  and Bi 
$p_z$-orbitals in the conduction band \cite{NatPhys.5.438}.  We use this 
characteristic to determine whether the material is behaving as trivial or 
topological insulator. Thus, by turning-off  the SOC, we observe 
that higher energy levels of the  valence band have higher contribution from 
the Se $pz$-orbitals, while the lower energy levels of the 
conduction band are dominated by Bi 
$pz$-orbitals. Meanwhile, by turning-on the SOC  we observe an inversion 
between valence and conduction band orbitals around the $\Gamma$--point, 
promoting the (band) parity  inversion. The same procedure was applied  to 
examine  if such a band inversion still occurs in py-Bi$_2$Se$_3$. 
By considering the pyridine concentration of 
1.64$\times$10$^{-2}$~mol./\AA$^2$ intercalated in Bi$_2$Se$_3$, we 
verify that when the SOC is not included the py-Bi$_2$Se$_3$ 
system is an insulator, where  the valence band maximum is dominated by Se 
$pz$-orbitals  and the conduction band minimum by Bi $pz$-orbitals 
[\Fig{bulk3mol}(a)],  as observed 
for the pristine Bi$_2$Se$_3$ bulk. In contrast, by turning SOC on we observe 
the lack of band inversion around the $\Gamma$--point [\Fig{bulk3mol}(b)], 
suggesting that pyridine molecules intercalated in Bi$_2$Se$_3$ bulk suppress 
its topological properties, namely, the py-Bi$_2$Se$_3$ system
is a trivial insulator. 
 
 This topological--trivial transition in the Bi$_2$Se$_3$ structure is possibly 
caused by the increase in the vdW gaps due to the pyridine intercalation. 
In order to verify such an assumption, we calculate the electronic 
band structure of Bi$_2$Se$_3$, but keeping the equilibrim geometry of the 
pyridine intercalated Bi$_2$Se$_3$ system. Our results are presented in 
Figs.\,\ref{bulk3mol}(c) and \ref{bulk3mol}(d), by turning-off and -on the SOC  
we observe that the band inversion at the $\Gamma$--point is  
still absent. 
 
\begin{figure}
  \centering
  \includegraphics[clip,width=8cm]{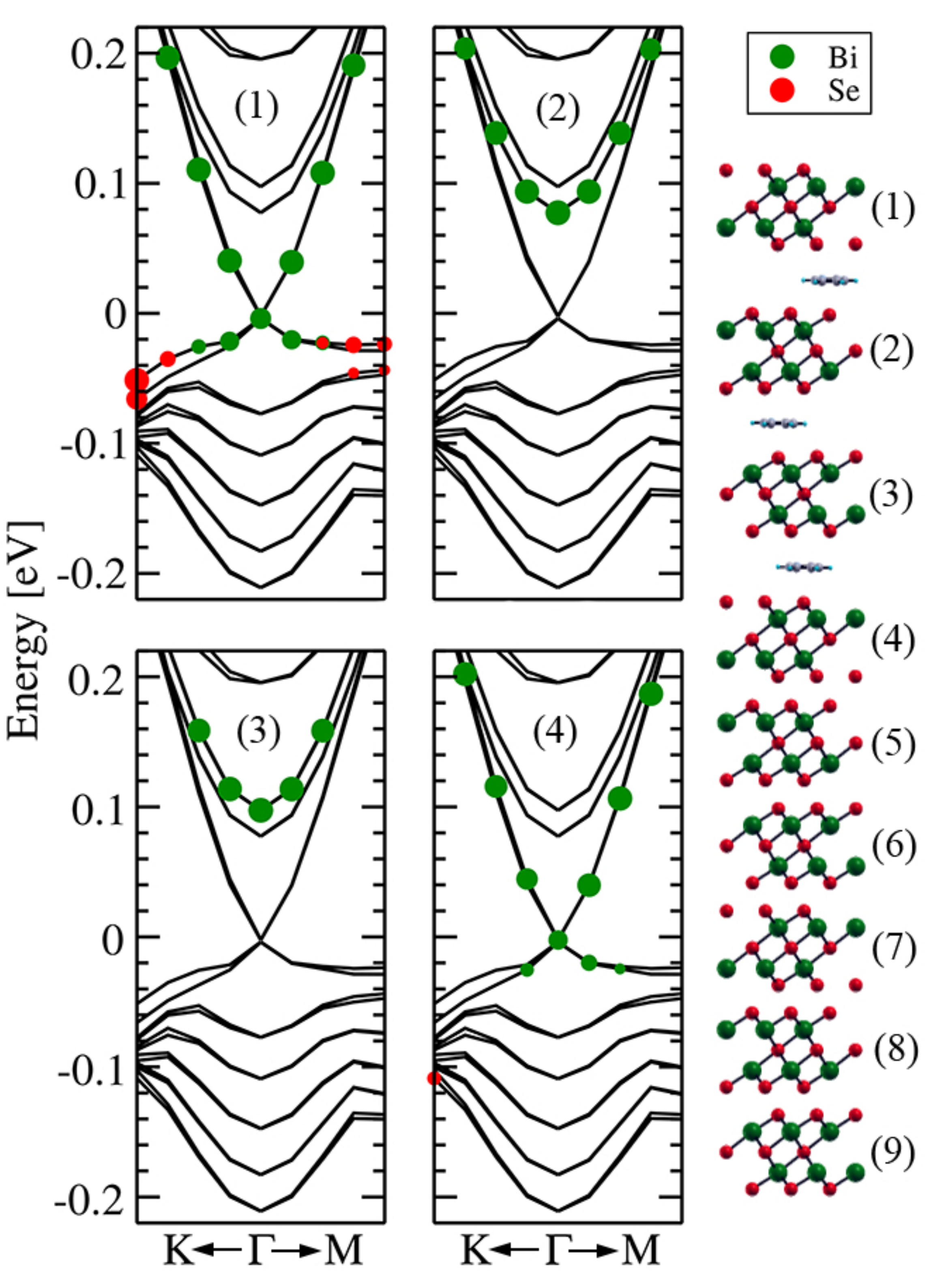}
  \caption{\label{hetbands} Each panel shows the band structure (full black lines)
   for the structure shown on the right side. The $pz$-orbital states is projected
   on the band structure for the first four QLs, where green circles represent Bi $pz$-orbitals
   and red circles Se $pz$-orbitals.}
\end{figure}

In experiments we would most likely observe regions with pyridine molecules 
intercalated between the vdW gaps and regions of pristine 
Bi$_2$Se$_3$, as a result we may have a trivial/topological  
insulator heterojunction composed by py-Bi$_2$Se$_3$ and Bi$_2$Se$_3$  
(py-Bi$_2$Se$_3$/Bi$_2$Se$_3$). To simulate such a heterojunction, we have 
considered a supercell  containing 9 QLs, where we have three 
QLs intercalated by the molecules and six consecutive pristine QLs,
(py-Bi$_2$Se$_3$)$_3$/(Bi$_2$Se$_3$)$_6$. As shown in 
\Fig{hetbands}, the pyridine molecules are intercalated between 
QL(1) and QL(4), characterizing the (py-Bi$_2$Se$_3$)$_3$ region, and 
QL(4)/QL(5)/.../QL(9)/QL(1) forms the pristine (Bi$_2$Se$_3$)$_6$ region.
Given that geometry, the trivial/topological interface are 
characterized 
by QL(1) and QL(4). Band structure calculations reveal the presence of metallic 
states forming a Dirac-like cone at the $\Gamma$-point, indicated 
by   solid black lines in \Fig{hetbands}. To localize these states we 
compute the $pz$-orbitals contribution from various QLs around the 
$\Gamma$-point, and project these states in the energy levels shown in the 
band structure\,\cite{parkPRL2010}. We find that the electronic 
states around the Dirac point are mainly attributed to the Bi 
$pz$-orbitals lying at the py-Bi$_2$Se$_3$/Bi$_2$Se$_3$ interface region,  
QL(1) and QL(4). In contrast, there are no electronic contribution to the 
metallic states from the  py-Bi$_2$Se$_3$ bulk region, QL(2) and QL(3). 
Somewhat similarly, the
electronic contributions to the Dirac-like cone coming from the bulk region 
of Bi$_2$Se$_3$ [QL(5)--QL(8)] are almost negligible. Those findings allow us 
to 
conclude that (py-Bi$_2$Se$_3$)$_{\rm m}$/(Bi$_2$Se$_3$)$_{\rm n}$  
heterostructures (or superlattices) give rise to topologically protected 
metallic states, embedded at the trivial/topological interface region. However, 
it is worth noting that the appearence of those topologically  protected sufaces 
states depends on the topological film (Bi$_2$Se$_3$) thickness (n). In 
Ref.\,\cite{yazyevPRL2010} the authors verified the presence of 
converged TSS for 3 QLs  of Bi$_2$Se$_3$. On 
the other hand, in a recent study, we verify the formation of TSSs  for an 
interlayer spacing, {\it i. e.} vdW gap, larger  than 
$\sim$5.5\,\AA\,\cite{seixasJAP2013}. Thus,  
we can infer  that it is possible to get topologically protected metallic 
channels by  considering just a single layer of py-Bi$_2$Se$_3$ embedded in 
Bi$_2$Se$_3$.

\begin{figure}
  \centering
  \includegraphics[clip,width=8cm]{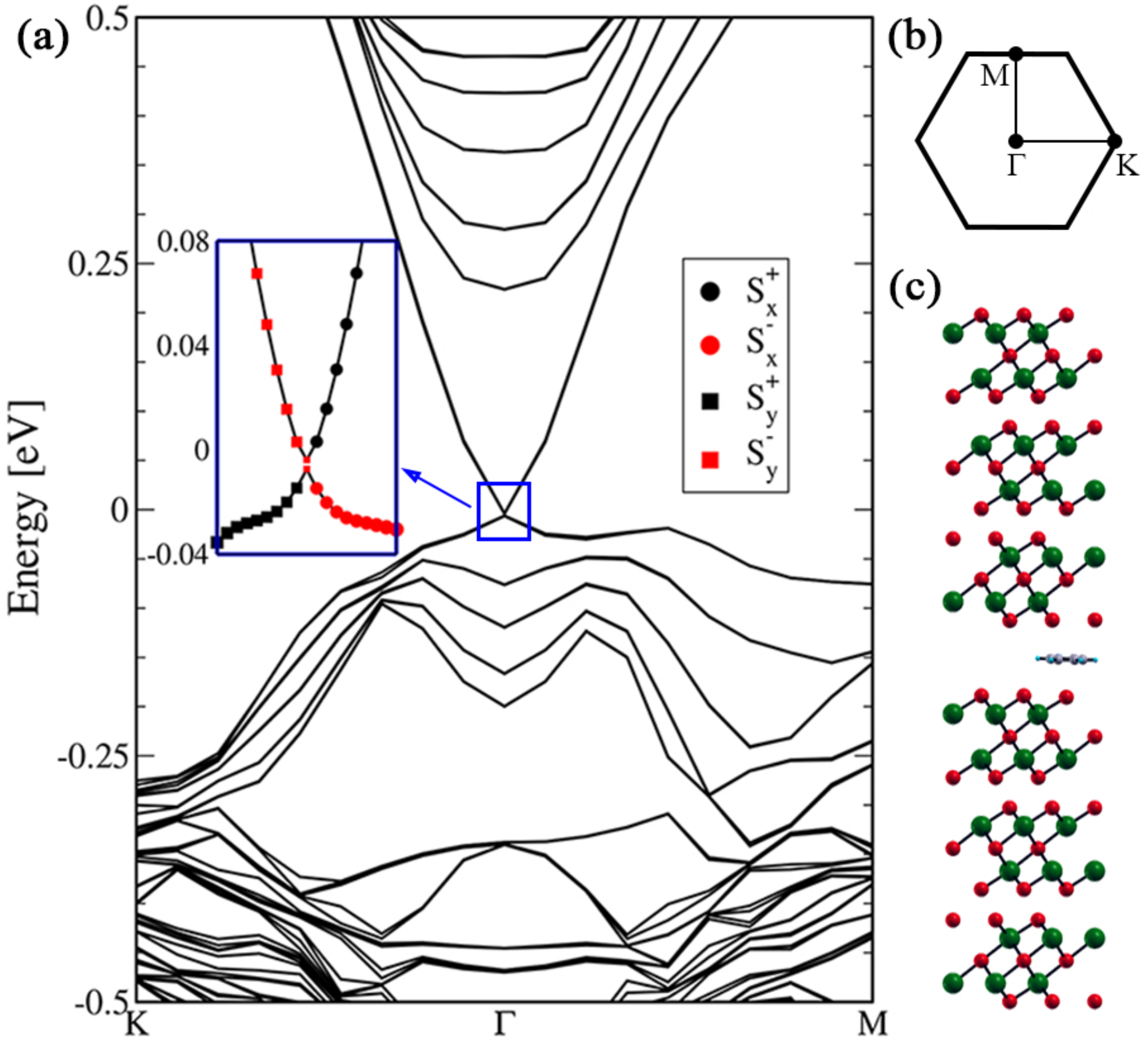}
  \caption{\label{6ql1mol} (a) Band structure for the system depicted in (c), along 
  the path shown in the projected surface 2D Brillouin zone represented in (b).   
  The inset in (a) shows the spin-texture inside the blue square 
  region around the $\Gamma$--point.}
\end{figure}

In order to verify the statement above,
we construct a supercell containing one molecule separated by six 
Bi$_2$Se$_6$   QLs, (py-Bi$_2$Se$_3$)$_{\rm 1}$/(Bi$_2$Se$_6$)$_5$,  
this structure is shown in \Fig{6ql1mol}~(c). 
Indeed, we find the TSSs 
forming a Dirac-like cone at the $\Gamma$-point, as shown in \Fig{6ql1mol}~(a).
Those states lie at the 
(py-Bi$_2$Se$_3$)$_{\rm 1}$/(Bi$_2$Se$_3$)$_5$ interface region. The 
spin-texture  of TSSs is constrained by the time reversal symmetry. Here we  
calculate the expected value 
of spin-polarization components ($\langle S_{n}({\bf k}) \rangle$) for the 
TSSs, of (py-Bi$_2$Se$_3$)$_{\rm 1}$/(Bi$_2$Se$_3$)$_5$, 
near the Dirac point. $\langle S_{n}({\bf k}) \rangle$ can be written as  
$\langle S_{n,\alpha}({\bf k}) \rangle = (\hbar/2)\langle\phi_n({\bf 
k})|\sigma_\alpha|\phi_n({\bf k})\rangle$, in cartesian coordinates ($\alpha = 
x, y, z$), where $\sigma$ represents the Pauli matrices, and $\phi_n({\bf k})$ 
the single particle Kohn-Sham eigenfunction, and $n$ 
represents the band index. We have considered $\phi_n({\bf 
k})$ for  wave  vectors ($\bf k$) along the $\Gamma$-M and 
$\Gamma$-K directions. As depicted in the inset of Fig.\,\ref{6ql1mol}(a),  for 
the electronic states parallel to the $\Gamma$-M direction, we find (i) positive 
(negative) values of $\langle 
S_{n,x}({\bf k}) \rangle$ for the occupied (empty) states, while (ii) $\langle 
S_{n,y}({\bf k}) \rangle$ = $\langle S_{n,z}({\bf k}) \rangle$ = 0; whereas for  
$\bf k$ parallel to the $\Gamma$-K direction, we find (iii) negative (positive) 
values of $\langle S_{n,y}({\bf k}) \rangle$ for the occupied (empty) states, 
(iv) $\langle S_{n,x}({\bf k}) \rangle$ = 0, and 
(v) $\langle S_{n,z}({\bf k}) \rangle$ values are negligible around the $\Gamma$-point. 
Such a picture of $\langle S_{n,\alpha}({\bf k}) \rangle$, for 
the TSSs lying at the (py-Bi$_2$Se$_3$)$_{\rm 1}$/(Bi$_2$Se$_3$)$_5$ 
interface, is practically the same as that obtained for the TSSs on
the Bi$_2$Se$_3$(111) surface\,\cite{yazyevPRL2010,abdallaPRB2013}.

\begin{figure}
  \centering
  \includegraphics[clip,width=8cm]{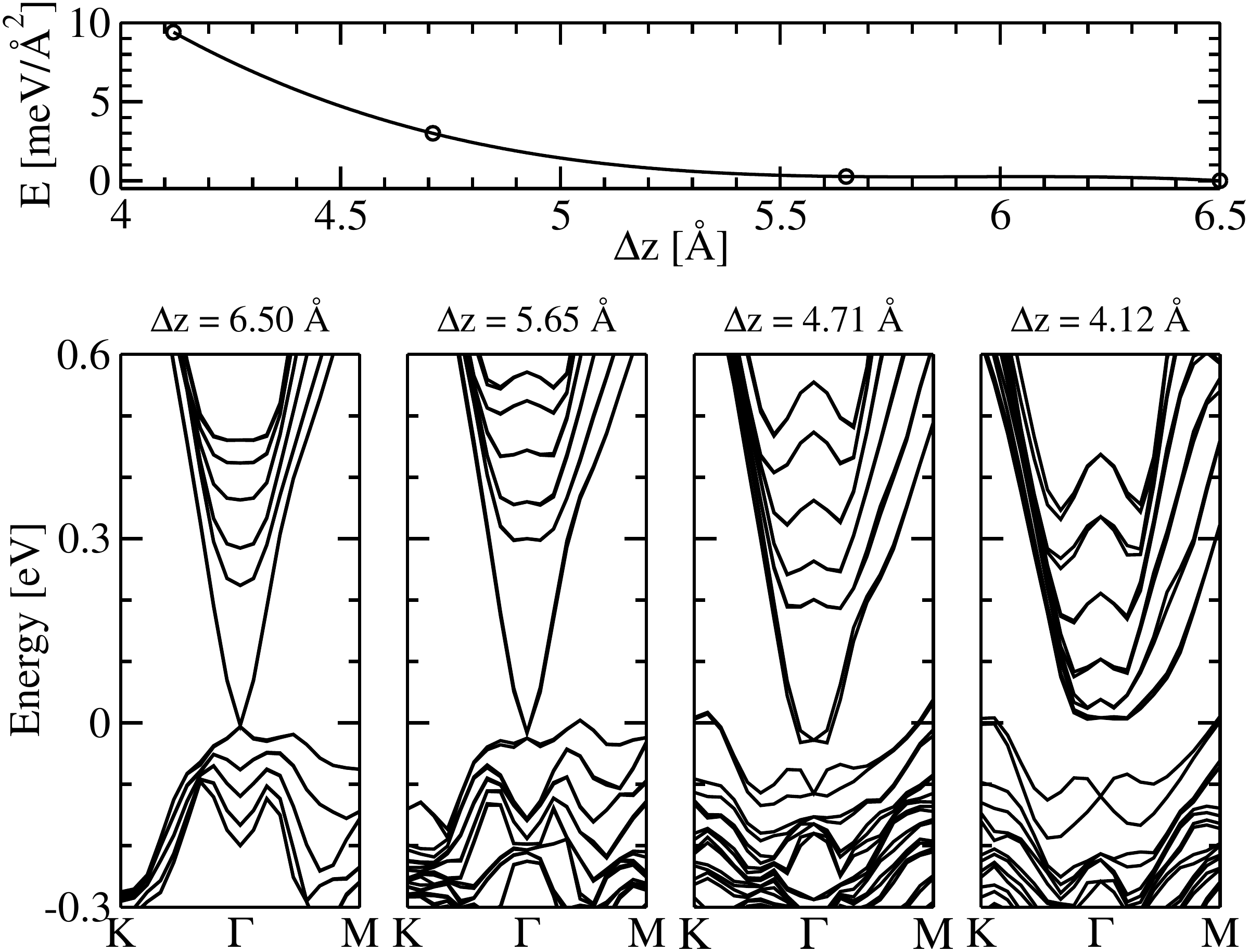}
  \caption{\label{6qlPres} \textit{Upper panel:} Energy 
  per area to compress the relaxed structure
  shown in \Fig{6ql1mol}~(c), reducing 
  the vdW gap between the QLs separated by the molecule.
  \textit{Lower panels:} Band structure for the various $\Delta z$ values.}
\end{figure}

We now propose a mechanism to gain control over the topologically protected 
metallic states in (py-Bi$_2$Se$_3$)$_{\rm m}$/(Bi$_2$Se$_3$)$_{\rm n}$. In 
particular, we considered the (py-Bi$_2$Se$_3$)$_{\rm 1}$/(Bi$_2$Se$_3$)$_{\rm 
5}$ system. We apply a compressive pressure directed along the $c$--axis, namely 
normal to the QLs, decreasing the separation distance in the vdW gap ($\Delta 
z$) from 6.50~\AA\ to 4.12~\AA, upon an external pressure of  $P\approx 
9.1$~GPa. This (pressure) value is comparable to the ones achieved 
experimentally for Bi$_2$Se$_3$ structures \cite{JPhysCondensMatter.24.035602, 
PhysRevLett.111.087001}. The energy difference between the relaxed and 
compressed (py-Bi$_2$Se$_3$)$_{\rm 1}$/(Bi$_2$Se$_3$)$_5$ is shown in the upper 
part of \Fig{6qlPres}. An energy of $\sim$~10~meV/\AA$^2$ needs to 
be added to compress the system from $\Delta z$ = 6.50 to 4.12~\AA. In the lower 
part of \Fig{6qlPres} we show the band structure for four values of $\Delta z$, 
starting from 6.50~\AA\ which has already been shown to have metallic states. 
For $\Delta z = 5.65$~\AA, the metallic states at the $\Gamma$-point are still 
present. Upon further  decrease of  $\Delta z$, the  metallic states start to 
move in the bulk band direction. For $\Delta z$ = 4.71~\AA, the Dirac point has 
been suppressed, as we find an energy gap at the $\Gamma$-point. The energy gap 
at the $\Gamma$-point is even larger for $\Delta z = 4.12$~\AA, washing out the 
TSSs at the (py-Bi$_2$Se$_3$)$_{\rm 1}$/(Bi$_2$Se$_3$)$_5$ interface region. By 
removing the pressure the system can reversibly return to its original geometry, 
and again the metallic states will be present. Thus, we can combine pressure 
application and pyridine molecules intercalation to the Bi$_2$Se$_3$  structure 
to turn the topologically protected metallic states on and off, 
going from an insulator to a semi-metallic material by taking advantage of the 
TI properties of Bi$_2$Se$_3$. The results presented above can be 
extended to any number of QLs intercalated by molecules, assuring that we have a 
 Bi$_2$Se$_3$/py-Bi$_2$Se$_3$ junction.

We have performed an {\it ab initio} study of pyridine molecules intercalated in 
the vdW gaps of Bi$_2$Se$_3$ (py-Bi$_2$Se$_3$). In py-Bi$_2$Se$_3$ the inter-QL 
distance increases, which turns the TI material into a trivial insulator. By 
considering  (py-Bi$_2$Se$_3$)$_{\rm m}$/(Bi$_2$Se$_3$)$_{\rm n}$ 
heterojunctions, we find a trivial/topological interface, characterized by the 
presence of topologically protected metallic states (forming a Dirac-like cone) 
embedded at the interface region of the heterostructure. Such metallic states 
can be present even for a single QL incorporated by pyridine molecules,   
(py-Bi$_2$Se$_3$)$_{\rm 1}$/(Bi$_2$Se$_3$)$_{\rm n}$. Lastly, we have shown the 
possibility to control the occurrence of such metallic states in 
(py-Bi$_2$Se$_3$)$_{\rm m}$/(Bi$_2$Se$_3$)$_{\rm n}$, upon an external 
compressive strain; turning-on and -off  those metallic states at the 
heterojunction interface.


\begin{center}
 
{\large\bf Acknowledgments}
 
\end{center}
 
This work was supported by the Brazilian Nanocarbon Institute of Science and
Technology (INCT/Nanocarbono), and the Brazilian agencies CNPq and FAPEMIG. The 
authors also acknowledge the computational support from CENAPAD/SP.



\begin{thebibliography}{25}
\bibitem [{rel()}]{relax}%
  \BibitemOpen
  \href@noop {} {}\bibinfo {note} {The atomic positions of the Bi$_2$Se$_3$ QLs
  and the intercalated molecules were fully relaxed.}\BibitemShut {Stop}%
\end{thebibliography}

%

\end{document}